\documentclass{article}
\usepackage{spconf,amsmath,graphicx}
\usepackage{espfig,graphicx}
\usepackage{enumerate}
\usepackage{multirow}
\usepackage{array}
\usepackage{color}
\usepackage{flushend}
\usepackage[numbers,sort&compress]{natbib}
\usepackage{subcaption}
\usepackage{graphicx,changepage}
\usepackage{multicol, blindtext}
\usepackage{calrsfs}
\usepackage{epstopdf}

\usepackage{tikz-cd}

\DeclareGraphicsExtensions{.eps}
\usepackage{amsmath,amssymb}
\usepackage{breqn}
\DeclareMathAlphabet{\pazocal}{OMS}{zplm}{m}{n}

\title{Two Channel Audio Zooming System for Smartphone}

\twoauthors
{A. khandelwal, E.B. Goud, Y. Chand, L. Kumar, S. Prasad}
	{
	 Electrical Engineering Department\\
	 Indian Institute of Technology, Delhi\\
	 anant.iitd.2085@gmail.com\\
	 \{ee172241/39,lkumar,sprasad\}@ee.iitd.ac.in\\
	} 
 {N. Agarwala, R. Singh}
	{
	 SRI Noida\\
	 ritesh.s7@samsung.com\\
	 n.agarwala@samsung.com
	} 
\begin{document}
\ninept
\maketitle
\begin{abstract}
In this paper,  two microphone based systems for audio zooming is proposed for the first time. The audio zooming application allows sound capture and enhancement from the front direction while attenuating interfering sources from all other directions. The complete audio zooming system utilizes beamforming based target extraction. In particular, Minimum Power Distortionless Response (MPDR) beamformer and Griffith Jim Beamformer (GJBF) are explored. This is followed by block thresholding for residual noise and interference suppression, and zooming effect creation. A number of simulation and real life experiments using Samsung smartphone (Samsung Galaxy A5) were conducted. Objective and subjective measures confirm the rich user experience.
\end{abstract}

%%%%%%%%%%%%%%%%%%%%%%%%%%%%%%%%%%%%%%%%%%%%%%%%%%%%%%%%%%%%%%%%%%%%%%%%%%%%%%%%
\vspace{-0.2cm}
\section{INTRODUCTION}
\begin{figure}[h]
\centering
\includegraphics[scale = 0.4]{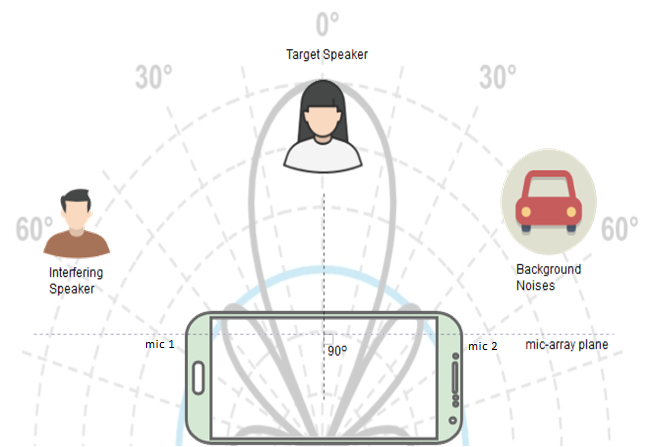} %dia.png
\caption{Two microphone smartphone based audio zooming: schematic depiction}
\label{prblm}
\end{figure}

Portable devices for communications like smartphones have become an inseparable part of life. The increasing dependency on the smartphones is due to numerous features they support, ranging from health and convenience to entertainment. One such useful feature being developed is audio zooming \cite{lee2016mobile, avendano2015audio} where sound from desired direction is enhanced while suppressing interferences from all other directions. This is desirable while trying to listen to a sound in the presence of one or more noise and interfering sources. Practical examples of such environments include that of a railway station, a stadium, classroom and market place. A pictorial depiction of the audio-zooming application for two microphone based smartphone is presented in Figure \ref{prblm}.
The evolution of compact device technology and computational power have resulted in use of multiple microphones in a smartphone to exploit the spatial diversity. Many smartphones today utilize two or more microphones. Apple iPhone-5\footnote{https://www.idownloadblog.com/2012/09/12/iphone-5-three-mics/} makes use of three microphones for beamforming and noise cancellation. Audio zooming has also been reported recently in some smartphones \footnote{https://www.youtube.com/watch?v=zzTUAcZ8FRQ,\\ https://www.youtube.com/watch?v=YNh4snIzmq4}. However, a significant improvement is required in the presence of severe noise, reverberation and multiple interferences. To the best of our knowledge, the only scientific publication for audio zooming in smartphone is \cite{duong2017audio} that utilizes MVDR beamforming with a linear array of four microphones.

\begin{figure*}[ht]
\centering
%\rule{\linewidth}{3cm}
\includegraphics[width=0.8\linewidth,height=6cm]{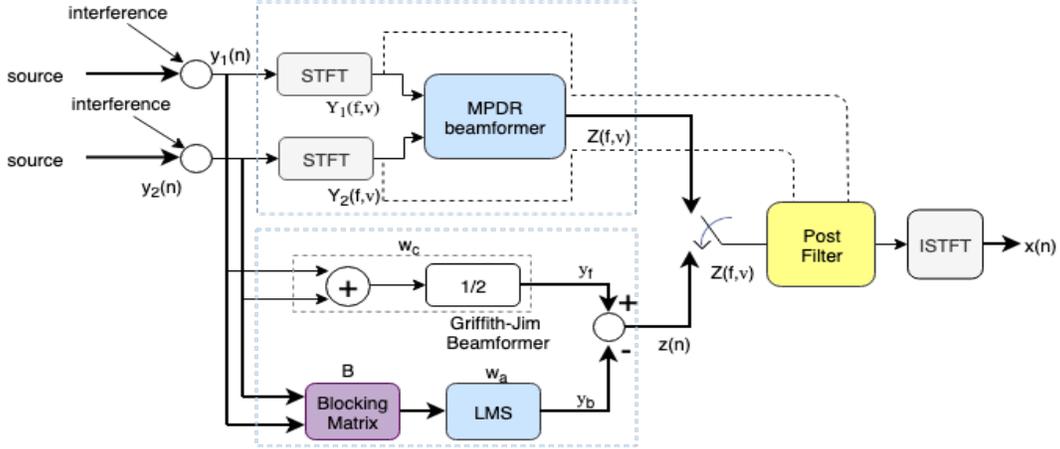}
\caption{Two stage audio zooming system using modified Griffith-Jim and MPDR beamformer}
\label{sys}
\end{figure*}

As most of the current smartphones have two microphones, the possibility of real time audio zooming with smartphone having two microphones is explored in this paper. The complete audio zooming system consists of two blocks as shown in Figure \ref{sys}. For the beamforming block, the simplest two channel frequency and time domain beamforming is investigated due to limited degree of freedom available. In particular, Minimum Power Distortionless Response (MPDR) beamformer \cite{caponJ} and Griffith Jim Beamformer (GJBF) \cite{griffiths1982alternative} are explored in frequency and time domain respectively. The two channel beamformer extracts the target source. This spatial filtering causes some suppression of the interference, but a significant residual interference may be present due to the use of only two microphones. A novel block-thresholding \cite{yu} based post-filtering is formulated  for creating the audio zooming effect. The additional novelty of the work is in exploration of two channel based audio zooming system deployable on smartphone. The filter length in proposed time domain GJBF can be estimated dynamically based on different environmental condition.
%%%%%%%%%%%%%%%%%%
%%%%%%%%%%%%%%%
%%%%%%%%%%%%%%%%%%%%%%%%%%%%%%%%%%%%
\vspace{-0.7cm}
\section{The Proposed Audio Zooming System}
We consider a smartphone with two identical and omni-directional microphones located at the top and the bottom. The target source to be acoustically zoomed in, is made incident on the smartphone normal to the plane containing the microphones, as shown in Figure \ref{prblm}. The sources incident from other directions are assumed to be interferences for the audio zooming application. The target and the interference are assumed to be in the far-field region. The proposed audio zooming systems consist of beamforming followed by post-processing. In particular, two audio zooming systems based on frequency and time domain beamforming, have been proposed and their performance have been analyzed . Motivation for using time domain based Griffith Jim Beamformer (GJBF) comes from the fact that it can control the level of interference at the output without the directional information of interferences \cite{griffiths1982alternative}. In the ensuing Section, two channel based MPDR beamformer is presented, followed by modified time domain GJBF. 
%%%%%%%%%%%%%%%%%%%%%%%%%
\subsection{Two Channel MPDR Beamformer}
A general wideband array data model can be written in STFT domain as  
\begin{equation} \label{fd_data_model}
\mathbf{Y}(f,\nu)= \mathbf{D}(\Psi,\nu)\mathbf{S}(f,\nu)+\mathbf{N}(f,\nu)
\end{equation}
where $\mathbf{Y}(f,\nu)$ is the received array signal, and $\mathbf{N}(f,\nu)$ is zero
mean, uncorrelated sensor noise. For $M$ microphones and $L$ sources, $\mathbf{D}(\Psi,f)$ is $M\times L$ array manifold given by 
\begin{equation}
\label{eq:94}
\mathbf{D}(\Psi,f) = [\mathbf{d}(\Psi_1,f), \mathbf{d}(\Psi_2,f), \dots,  \mathbf{d}(\Psi_L,f)]
\end{equation}
where $\mathbf{d}(\Psi_l,f)$ represents the steering vector for $l^{th}$ source given by 
\begin{equation}
\label{eq:95}
\mathbf{d}({\Psi_l,f})= \begin{bmatrix}e^{-j2\pi f\tau_{1l}} & e^{-j2\pi f\tau_{2l}}& \dots & e^{-j2\pi f\tau_{Ml}}\end{bmatrix}^T
\end{equation}
$\Psi_l = (\theta_l,\phi_l)$ is the incident direction of the $l^{th}$ source. 
Here $\tau_{ml}$ represents the time delay of arrival of the $l^{th}$ signal at the $m^{th}$ microphone with respect to a reference microphone. MPDR beamforming problem is equivalent to minimizing the output power with a distortionless response in the target direction given as
\begin{equation}
    \underset{\mathbf{W}}{min}\hspace{0.1cm}\mathbf{W}^H \mathbf{R_Y}\mathbf{W} \hspace{0.3cm}subject \hspace{0.1cm}to \hspace{0.3cm}\mathbf{W}^H \mathbf{d}({\Psi_l,f}) = 1
\end{equation}
The solution to the above optimization problem under diagonal loading is given by 
\begin{equation}
\mathbf{\hat{W}}_{(\Psi_l,f)}=\frac{(\mathbf{\hat R}_{{Y},f} + \alpha\mathbf{I})^{-1}\mathbf{d}(\Psi_d,f)}{\mathbf{d}^H(\Psi_d,f)(\mathbf{\hat R}_{{Y},f} + \alpha \mathbf{I})^{-1}\mathbf{d}(\Psi_d,f)}
\end{equation}
where $\alpha$ is the diagonal loading factor. The co-variance matrix is estimated as
\begin{equation}
    R_{Y,f} = \frac{1}{K} \sum_{k = 0 }^{k=K} Y(f,k) *Y^{*}(f,k)
\end{equation}
where K is total number of time frames.
%%%%%%%%%%%%%%%%%%%%%%%%%%%%%%%%%%%%
\subsection{Modified Griffiths-Jim Adaptive Beamformer}
Audio zooming application is additionally, explored using two channel time domain beamformer.
The target is assumed to be incident from broadside, resulting in identical delays at the two microphones. The $n^{th}$ snapshot of the received signal at the $m^{th}$ microphone is written as 
\begin{equation}
    y_{m}(n) = s(n) + n_{m}(n), \textrm{  } n = \{0, 1, \cdots, N_s-1\}, m = \{1, 2\}
    \label{dm1}
\end{equation}
where $s(n)$ is the target signal to be zoomed in, and $n_{m}(n)$ is the total noise and interferences. 

As the target signal undergoes identical delays at the two microphones, no phase adjustment is required herein when compared to the original GJBF \cite{griffiths1982alternative}. The constrained weight $w_c=\frac{1}{2}[1, 1]^T$ for the target will result in simple addition of the two channel signal providing signal plus interference $y_f$ in the upper branch of GJBF in Figure \ref{sys}. The blocking matrix $B=[1,-1]^T$ subtracts the two channel data thus it does not allow the signal from the constraint direction in the lower branch of GJBF. The adaptive weight $w_a$ in the lower branch is chosen to estimate the signal at the output of $w_c$ as a linear combination of the data at the output of the blocking matrix $B$. As blocking matrix does not allow the signal from the constraint direction, the signal estimated by $w_a$ is the interference close to interference present at the output of $w_c$. 
%%%%%%%
\begin{figure*}[h]
\centering
\begin{minipage}{0.5\textwidth}
	\includegraphics[width=\linewidth,height = 2cm]{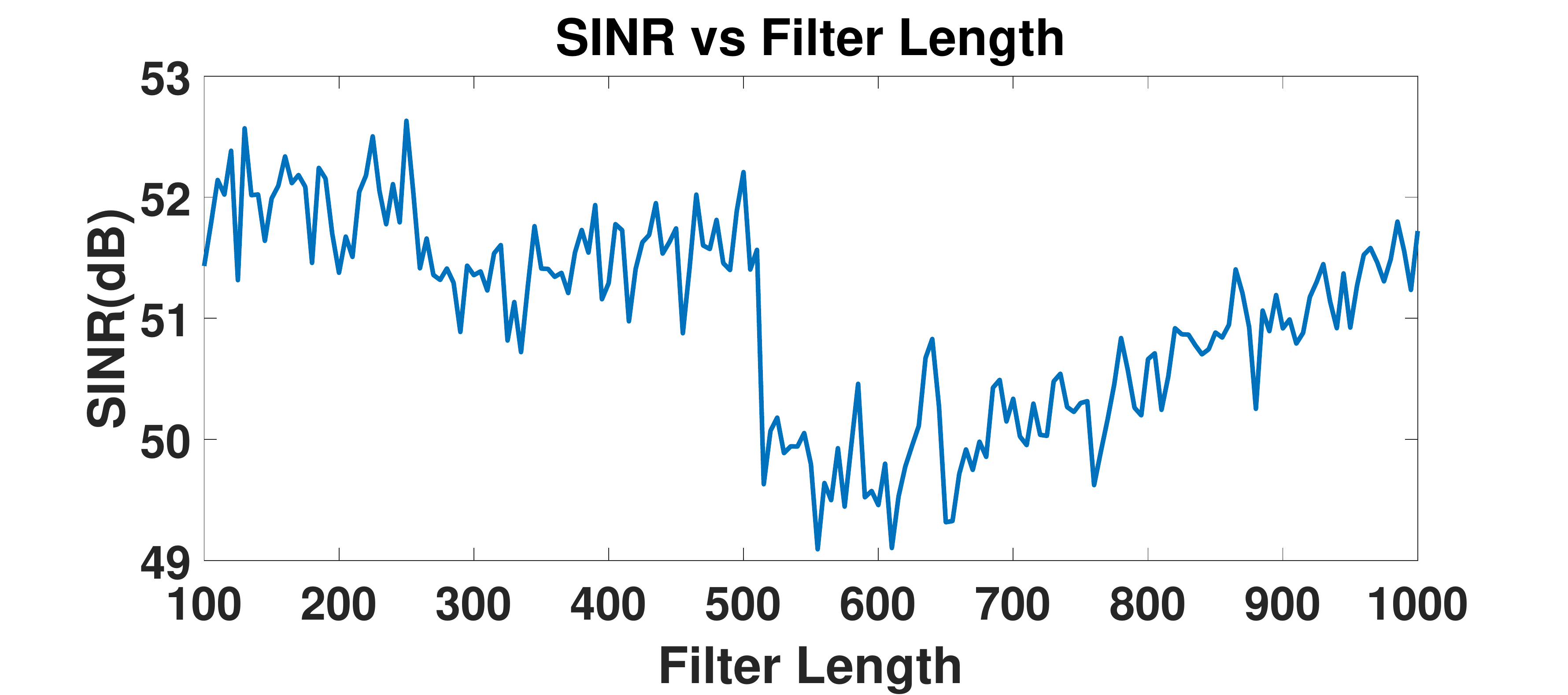}
	\caption*{(a) Anechoic recording.}
\end{minipage}%
\begin{minipage}{0.5\textwidth}
	\centering
\includegraphics[width=\linewidth,height=2cm ]{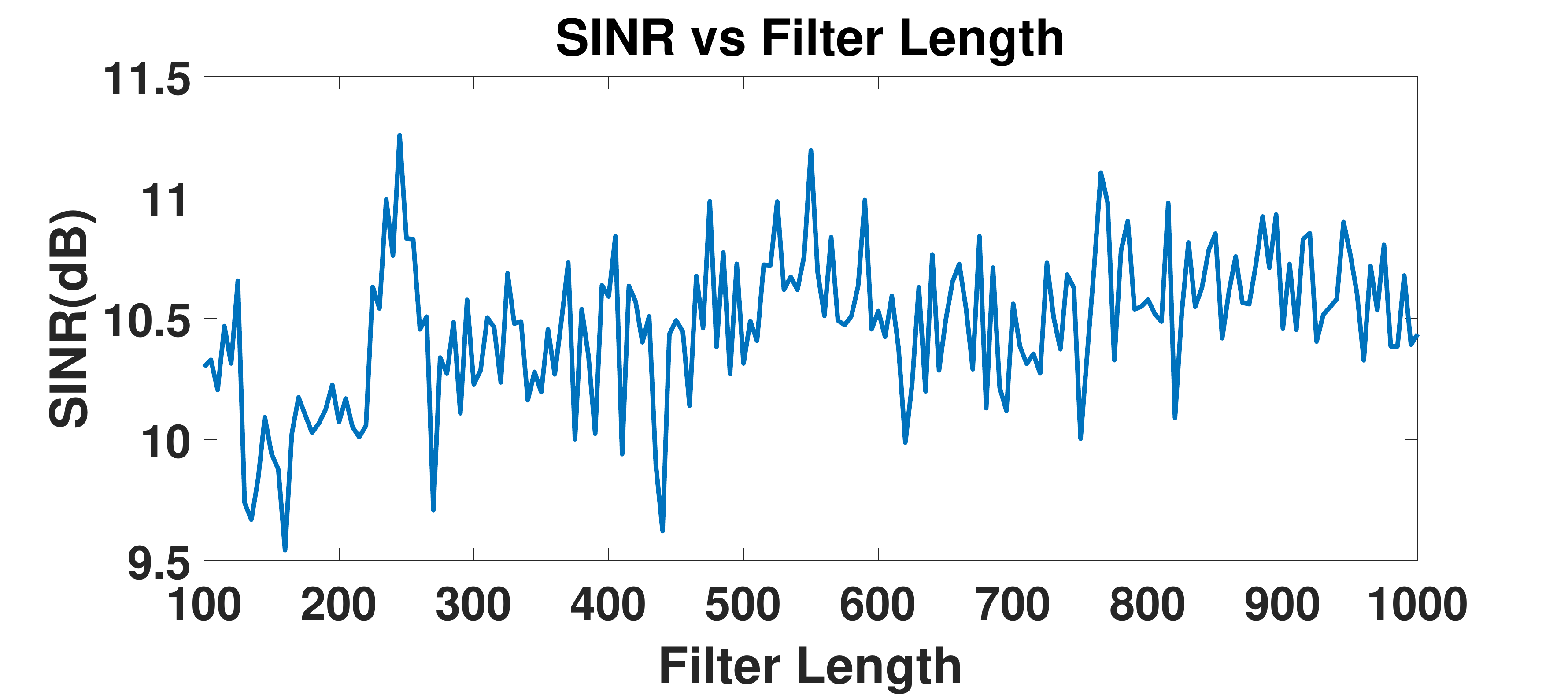}
\caption*{(b) A Meeting room recording.}
\end{minipage}
\caption{Dynamic LMS Filter Length Estimation (a) The SINR is max at filter length of 250 (b) The SINR is maximum for filter length 245.}
\label{dynm}
\end{figure*} 
%%%%%%%%%%%%%%%%%%%%%

The overall output of the two microphone GJBF is 
\begin{equation}
z(n)=y_f(n)-y_b(n), 
\end{equation}
where $y_f$ has the target signal, noise and interference with response determined by $w_c^H$, and $y_b$ has only the noise and interference. The filter weights $w_a$ is found based on minimizing the power contained in $z(n)$. The power will be minimum when $y_b(n)$ closely models the interference and noise present in the $y_f(n)$. In particular, Frequency Domain Adaptive Filter (FDAF) approach \cite{mansour1982unconstrained} using overlap-save method has been utilized for computing $w_a$. The FDAF approach has logarithmic complexity when compared to polynomial complexity for the classical LMS as utilized in the original GJBF.
%%%%%%%%%%%%%%%%%%%%%%%%%%%%%%%%%%%%%%%%%%%%%%%
\subsubsection{LMS Filter Length}
The GJ beamformed output quality depends on the length of LMS filter. The filter length depends on the environmental conditions. The length of the filter can be dynamically decided based on the highest SINR in GJBF output. The SINR is computed in Short Time Fourier Transform (STFT) domain as $\frac{|Z(f,\nu)|^2 - \sigma^2(f,\nu)}{\sigma^2(f,\nu)}$, where the estimation of residual noise variance is detailed in Section \ref{blockth}. SINR at GJBF output is plotted in Figure \ref{dynm} with the LMS filter length for anechoic chamber and reverberant room recording. It is to note that the optimum filter length for anechoic set-up is $250$ while that of reverberant room is $245$.
\subsection{Adaptive Block Thresholding based Post-Filtering for Audio Zooming} \label{blockth}
In this Section, adaptive block thresholding based post-filtering is proposed for Audio Zooming effect creation. The estimate obtained by adaptive beamforming recovers the target with some amount of noise and interference. Additionally, the beamformed output might show transient or tonal behavior, or combination of the two. The residual interference along with transient and tonal behavior is now simply treated as single additive interference term for further processing. Mathematically, the beamformed output can now be written in STFT domain as
\begin{equation}
Z(f,\nu) = S(f,\nu)+I(f,\nu)
\label{bfop}
\end{equation}
where $S(f,\nu)$ is STFT  coefficient of desired signal and $I(f,\nu)$ is STFT coefficient of residual noise and interference. The total variance of such additive interference can be computed as
\begin{equation}
\hat{\sigma}^2 (f,\nu)= \frac{1}{2} \{(Y_{1}(f,\nu) - Z(f,\nu))^{2} +(Y_{2}(f,\nu) - Z(f,\nu))^{2}\}
\end{equation}
where $Y_1(f,\nu)$, $Y_2(f,\nu)$ are the STFT coefficients of microphone 1 and 2 respectively. This is the best residual interference variance that can be estimated considering the beamformed output is close to the target signal. This estimate works well with practical scenarios. More accurate estimation can provide better result. It is to be noted that because of limited degrees of freedom (two, utilized both) in the spatial domain, post-filtering is now attempted in frequency domain. For this purpose, the output of the either beamformer is considered in short-time Fourier transform (STFT) domain, with suitably chosen block sizes in the time and frequency, as detailed below.

The time-frequency coefficients are modified by multiplying each of them by an attenuation factor to attenuate the interference dominated components as
\begin{equation}
\hat{S}(f,\nu) = a(f,\nu)*Z(f,\nu)
\end{equation}
and creating the zooming effect.
%%%%%%%%%%%%%%%
\begin{figure}
\centering
	\includegraphics[width=\linewidth]{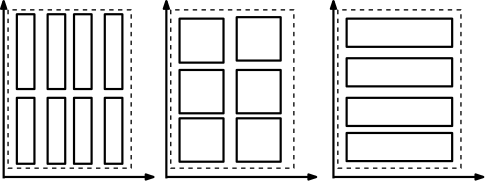}
	\caption{Example of dividing a macro-block into sub-blocks}
	\label{fug42}
\end{figure}
%%%%%%%%%%%%%%%%%%%%%%
The attenuation factor $a(f,\nu)$ depends upon the values $Z(f',\nu')$ for all $[f',\nu']$ in the neighborhood of $[f,\nu]$. The signal estimate $\hat{S(f,\nu)}$ is computed from the noisy data $Z(f,\nu)$ with a constant attenuation factor $a_{i}$ over the sub-block $ B_{i}$ as
\begin{equation}
\hat{S}(f,\nu) = a_{i}\hspace{0.1cm}  Z(f,\nu) \textrm{ }\forall (f,\nu)\in B_{i}
\end{equation}
Selection of the block is done as follows. The entire STFT matrix $\mathbf{Z}$ is divided into macro-blocks of size ${P\times Q}$. Each macro-block is further divided into sub-blocks of size $2^{H-v} \times 2^v$ where $2^H=L$ and $2^V=W$ and $v \in \{0,1.....H\}$. Various ways of dividing a macro-block into sub-blocks is shown in Figure \ref{fug42}. A Threshold block SNR is chosen to differentiate higher and lower block SNR values. Out of the various structure of sub-blocks, the one having highest SNR is chosen.

The mean square error can be written as
\begin{equation}
r = E(| \hat{S} - S|^{2})\leq \frac{1}{A} \sum^{I}_{i=1} \sum_{f,\nu \in B_{i}} E \{(a_{i} \hspace{0.1cm}  Z(f,\nu) - S(f,\nu) )^{2} \}
\end{equation}
%%%%%%%%%%%%%%%%%%%%%%%%%%%%%%%%%%%%%%%%%%%%%%%%%%%%%%%%%%%%%%%%%%%%%%%%%%%%%%%%%%%%%%
%%%%%%%%%%%%%%%%%%%%%%%%%%%%%%%%%%%%%%%%%%%%%%%%%%%%%%%%%%%%%%%%%%%%%%%%%%%%%%%%%%%
The error can be minimized by choosing \cite{attenuation}
\begin{equation}
a_{i} = (1 -\frac{1}{\zeta_{i}+1})_{+}\footnote{a = $(x)_{+}$ $\implies$ $a=x$ if $x > 0$ else $a=0$}  
\end{equation}
where $\zeta_{i} =\frac{\bar{S}^{2}}{\bar{\sigma}^{2}}$ is the average apriori SNR in the sub-block $ B_{i} $, that is computed from  
\begin{equation}
\bar{S}^{2} = \frac{1}{B^{0}_{i}}\sum_{f,\nu \in B_{i}} S^{2}(f,\nu) \textrm{and }, \hspace{0.5cm} \bar{\sigma}^{2} = \frac{1}{B_{i}^{0}} \sum_{f,\nu \in B_{i}} \sigma^{2}(f,\nu)
\label{avrg}
\end{equation}
Here $B_{i}^{0}$ is number of coefficients in the sub-block. As $S(f,\nu)$ is unknown, the apriori SNR  ${\zeta}_{i}$ can be computed alternatively using 
\begin{equation}
\hat{\zeta}_{i} =  \frac{\bar{Z_{i}}^{2} }{\bar{\sigma_{i}}^{2}} -1
\end{equation}
that can be derived from (\ref{bfop}). $\bar{Z_{i}}^{2}$ can be computed as in (\ref{avrg}) using the beamformed signal. 

It is to be noted that the selection of $a_i$ ensures that the target signal is enhanced corresponding to the high SNR sub-blocks while suppressing the low-SNRs sub-blocks. This results in the required audio-zooming effect. 
%%%%%%%%%%%%%%%%%%%%%%%%%%%%%%%%%%%%%%%%%%%%
%%%%%%%%%%%%%%%
\begin{figure}
\centering
\includegraphics[width= 6cm, height=0.3\linewidth]{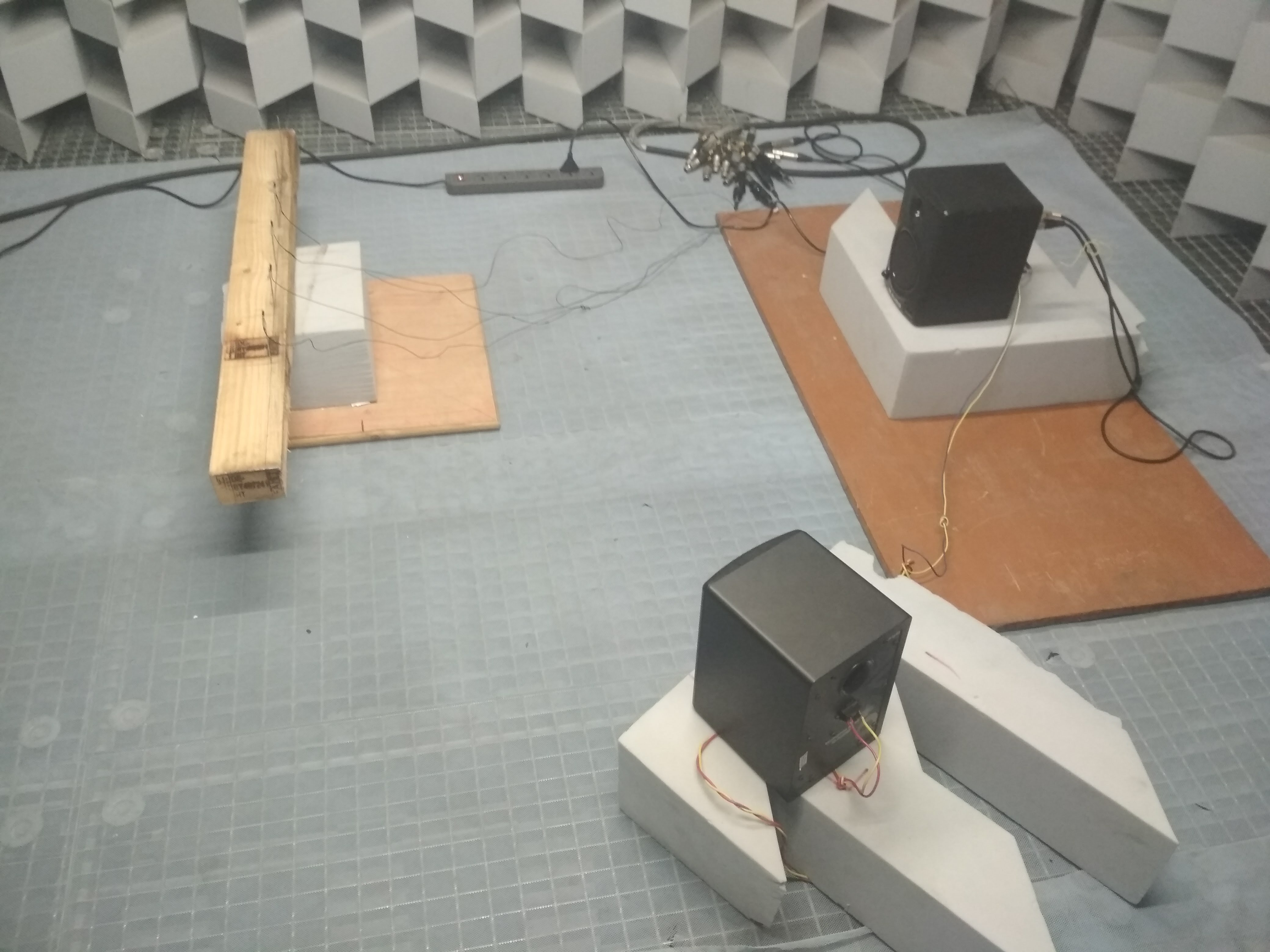}
\caption{Anechoic Chamber Recording Setup}
\label{ana}
\end{figure}
%%%%%%%%%%%%%%%%%%%%%%
\section{Performance evaluation}
The performance evaluation of the proposed audio zooming systems is presented herein using simulation and real data experiments. Four objective measures that include Mean Square Error(MSE) \cite{emiya2011subjective}, Output SINR (OSINR) \cite{emiya2011subjective}, Perceptual Evaluation of Speech Quality (PESQ) \cite{rix2001perceptual} and Short-time Objective Intelligibility (STOI) \cite{taal2010short} are utilized for the simulation experiments. Experiments were additionally conducted on real data recorded in anechoic chamber and in the field. The Mean Opinion Score (MOS) \cite{viswanathan2005measuring} measure is utilized to evaluate the performance of experiments with real recorded data. Non-refference PESQ score is also given to evaluate the speech intelligibility. The proposed time and frequency domain audio zooming systems are compared with RMVDR based system \cite{duong2017audio} for two microphones. 
\vspace{-0.3cm} 
\subsection{Simulation Experiments}
Two microphones separated by $10$ cm were taken for the simulation experiments. The objective parameters were computed for fifty target speech files from TIMIT database \cite{zue1990speech}. The target source was taken at azimuth $90^\circ$, while the interference was assumed to be at $60^\circ$.  Signal-to-Interference Ratio (SIR) was taken to be $0$dB. The experiments were performed in reverberant condition. For reverberant condition, the reverberation time was taken to be $100$ms. The results are presented in Table \ref{objective_eval}. It is to be noted that the proposed MPDR beamforming based system outperforms the other methods. The performance of GJBF based system is comparable to RMVDR based system.    
%\begin{figure}[ht]
%    \centering
%\includegraphics[width=0.8\linewidth, height=0.4\linewidth]{IMG_20180125_154916.jpg}
%    \caption{Anechoic Chamber Recording Setup}
%    \label{ana}
%\end{figure} 
%%%%%%%%%%
%%%%%%%%%%%%%%%%%%%%%
\begin{table}[ht]
\centering
\begin{tabular}{|p{1.5cm}|l|l|l|l|}
\hline
\textbf{Zooming System}    & \textbf{OSINR(dB)} & \textbf{PESQ} & \textbf{STOI} & \textbf{MSE(dB)} \\ \hline
RMVDR          & 10.56          & 3.093        & 0.8935          & -25.83       \\ \hline

Griffith Jim  &      9.35       & 3.0572        &   0.8655   & -23.56       \\ \hline

MPDR                    &  
13.85          &  3.203        &      0.9175    & -33.453       \\ \hline

\end{tabular}
\caption{Objective Evaluation of proposed audio zooming systems}
\label{objective_eval}
\end{table}
%%%%%%%%%%%%
\vspace{-0.8cm}
\subsection{Experiments on Real Data}
The data recording was done was in anechoic chamber and open field. Subjective listening tests were conducted. Mean Opinion Score (MOS) measure is presented to evaluate the proposed audio zooming system. Fifteen subjects in the age group of $20-25$ years were invited for listening the zoomed audio. The subjects listened the mixed received signal and the output. The rating was given for the quality of the output audio on the scale of $0$ to $5$ \cite{citation-0}.
\subsubsection{Microphone Array Recordings in Anechoic Chamber} 
A uniform linear microphone array was utilized for recording in anechoic chamber with inter-element spacing as $4.5$cm. The array consists of Sennheiser HSP 2 microphones. The target speaker was placed at $90^\circ$ azimuth. The interference was kept at $40^\circ$ azimuth.Various kind of interference were selected as shown in table 2. Data of two channel separated by $9$cm was utilized for evaluating the audio zooming systems. The MOS measures being close to or more than $4$ shows a good perception of the output.
% Please add the following required packages to your document preamble:
% \usepackage{multirow}
\begin{table}[h!]
\begin{tabular}{|c|c|c|c|}
\hline
\multirow{2}{*}{\textbf{Zooming Technique}} & \multirow{2}{*}{\textbf{Interference}} & \multicolumn{2}{c|}{\textbf{MOS}}                                                                                                               \\ \cline{3-4} 
                                            &                                        & \textit{\begin{tabular}[c]{@{}l@{}}Interference\\ suppression\end{tabular}} & \textit{\begin{tabular}[c]{@{}l@{}}Speech\\ quality\end{tabular}} \\ \hline
\multirow{4}{*}{RMVDR}                       & Train                                  & 3.25                                                                        & 4                                                                 \\ \cline{2-4} 
                                            & Vacuum                                & 3.5                                                                         & 4                                                                 \\ \cline{2-4} 
                                            & Speech                                 & 3.25                                                                        & 3.75                                                              \\ \cline{2-4} 
                                            & Sonic                                  & 3.25                                                                        & 4                                                                 \\ \hline
\multirow{4}{*}{Griffith-Jim}               & Train                                  & 4.16                                                                       & 3.75                                                              \\ \cline{2-4} 
                                            & Vacuum                                & 4.25                                                               & 4                                                                 \\ \cline{2-4} 
                                            & Speech                                 & 4                                                                           & 3.75                                                              \\ \cline{2-4} 
                                            & Sonic                                  & 4.16                                                                       & 3.85                                                              \\ \hline
\multirow{4}{*}{MPDR}                       & Train                                  & 4                                                                           & 4.25                                                    \\ \cline{2-4} 
                                            & Vacuum                               & 4.25                                                             & 4.16                                                             \\ \cline{2-4} 
                                            & Speech                                 & 4.16                                                                       & 4                                                                 \\ \cline{2-4} 
                                            & Sonic                                  & 4.16                                                                      & 4.25                                                    \\ \hline
\end{tabular}
\label{PE_ane}
\caption{Performance evaluation for anechoic chamber recording}
\end{table}
%%%%%%%%%%%%%%%%%%%%%%%%%%%%%%%%%%%%%%
\subsubsection{Smartphone Recordings in Open Ground}
As the application target is smartphone, we used two channel recording from Samsung Galaxy A5(2017) smartphone. The recording was performed in a open ground scenarios. The Two orators were located at $90^\circ$ and $45^\circ$ to the microphone array. MOS and non-reference PESQ (NR-PESQ) \cite{nr} measures are given in Table \ref{rev}.  High MOS and NR-PESQ measures shows the practical applicability of the proposed systems. The more field recording results are made available online at http://web.iitd.ac.in/$\sim$lalank/msp/demo.html for reviewers to evaluate. 

\begin{table}[h!]
\begin{tabular}{|c|c|c|c|}
\hline
\multirow{2}{*}{\textbf{Zooming Technique}} & \multicolumn{2}{c|}{\textbf{MOS}}                                                                                                               & \multirow{2}{*}{\textbf{NR-PESQ}} \\ \cline{2-3}
                                            & \textit{\begin{tabular}[c]{@{}l@{}}\textbf{Interference}\\ \textbf{suppression}\end{tabular}} & \textit{\begin{tabular}[c]{@{}l@{}}\textbf{Speech}\\ \textbf{quality}\end{tabular}} &                                   \\ \hline
RMVDR                                        & 3.5                                                                         & 4                                                                 & 2.835                             \\ \hline
Griffith-Jim                                & 4.08                                                                        & 3.66                                                              & 2.756                             \\ \hline
MPDR                                        & 4.25                                                             & 4.166                                                    & 3.021                    \\ \hline
\end{tabular}
\caption{Evaluation for open ground recording}
\label{rev}
\end{table}
\vspace{-0.5cm}
\section{CONCLUSIONS}
In this paper, two channel audio zooming systems are proposed for smartphone for the first time. Two channel time and frequency domain beamforming based target extraction is explored. A novel block-thresholding based post-filtering is utilized for creating the audio zooming effect. The proposed MPDR and GJ beamformer based systems are compared with two channel RMVDR based system. The MPDR based audio zooming system outperforms while performance of GJBF based system is comparable with RMVDR. The proposed systems are tested on smartphones for the expected result. MPDR based system can be utilized for smartphones with more than two microphones with better output. A mobile app is being developed for the same. 
%and Griffith-jim systems performed well as compared with MVDR based system proposed in \cite{mvdr} in all the environments.
Subjective and objective measures suggest rich user experience.
%\section*{ACKNOWLEDGMENT}
\bibliographystyle{IEEEbib}
\bibliography{strings,ref}

\end{document}